\documentclass[letterpaper, 10 pt, conference]{ieeeconf}  

\IEEEoverridecommandlockouts                              

\usepackage{amsmath}
\usepackage{multirow}
\usepackage{graphicx}
\graphicspath{{figs/}}
\usepackage{caption}
\usepackage{subcaption}
\usepackage{booktabs}
\usepackage{xcolor} 

\title{\LARGE \bf
Optimizing BCI Rehabilitation Protocols for Stroke: Exploring Task Design and Training Duration
}

\author{Aniana Cruz$^{1}$ Marko Kuzmanoski$^{2}$ Gabriel Pires$^{3}$ 
\thanks{*This work has been supported by Fundação para a Ciência e a Tecnologia (FCT) under projects 2023.13809.PEX (DOI https://doi.org/10.54499/2023.13809.PEX), BCI4ALL (COMPETE2030-FEDER-00842800, 2023.17977.ICDT) and  UID/00048 - Instituto de Sistemas e Robótica - Coimbra (ISR-UC).}
\thanks{$^{1}$Aniana Cruz is with the Institute of Systems and Robotics, University of Coimbra, Portugal
        {\tt\small anianabrito@isr.uc.pt}}%
\thanks{$^{2}$Marko Kuzmanoski is with the Institute of Systems and Robotics, University of Coimbra, Portugal, and with University of Ljubljana, Slovenia
        {\tt\small mk0903@student.uni-lj.si}}%
\thanks{$^{3}$Gabriel Pires is with the Institute of Systems and Robotics, University of Coimbra, Portugal, and also with the Polytechnic Institute of Tomar, Portugal
        {\tt\small gpires@isr.uc.pt}}%
}

\begin{document}

\maketitle
\thispagestyle{empty}
\pagestyle{empty}

\begin{abstract}
Stroke is a leading cause of long-term disability and the second most common cause of death worldwide. Although acute treatments have advanced, recovery remains challenging and limited. Brain-computer interfaces (BCIs) have emerged as a promising tool for post-stroke rehabilitation by promoting neuroplasticity. However, clinical outcomes remain variable, and optimal protocols have yet to be established. This study explores strategies to optimize BCI-based rehabilitation by comparing motor imagery of affected hand movement versus rest, instead of the conventional left-versus-right motor imagery. This alternative aims to simplify the task and address the weak contralateral activation commonly observed in stroke patients. Two datasets, one from healthy individuals and one from stroke patients, were used to evaluate the proposed approach. The results showed improved performance using both FBCSP and EEGNet. Additionally, we investigated the impact of session duration and found that shorter training sessions produced better BCI performance than longer sessions.


\end{abstract}

\section{INTRODUCTION}

According to the World Health Organization from 1990 to 2019, the number of strokes increased dramatically by about 70\%, with more than 12.2 million new strokes each year. The estimated global cost of stroke is over US$721$ billion \cite{feigin2022world}. Post-stroke survivors suffer a significant loss of neuronal functions, mainly in the sensorimotor area. Therefore, in some cases, they have severe motor impairments and tend to lose a significant degree of autonomy in daily tasks, making them very dependent on others. Motor impairment in elderly individuals is up to 80\% and only about 20\% experience partial recovery \cite{emerson2018combined}. Despite advances in acute stroke treatment, functional recovery remains limited, highlighting the need for novel rehabilitation strategies.
Motor imagery (MI) therapy is one of the conventional approaches used in upper limb rehabilitation.
In motor imagery therapy, patients are instructed to mentally simulate movement of the affected limb without actual execution. However, there is no objective way to confirm whether MI is performed correctly. Brain-computer interfaces (BCIs), which decode users’ intentions from brain signals to control external devices with real-time feedback, offer a promising solution \cite{wolpaw2013brain}. By detecting motor intention, BCIs provide an objective measure of MI quality and enable feedback only when correct motor imagery is identified.
This strategy aligns with a fundamental principle of neurofeedback: rewards should only be given when patients perform tasks that deserve a reward \cite{neuper2014b}.
Several studies have shown that MI can activate the same neural networks involved in movement execution \cite{baniqued2021brain}. This evidence supports its relevance for rehabilitation, and when integrated into a BCI system, enables closed-loop operation \cite{de2020effects}. Motor imagery triggers event-related (de)synchronization (ERD/ERS) of sensorimotor rhythms (SMR). When combined with visual or proprioceptive neurofeedback, it may enable patients to learn how to voluntarily modulate SMR activity, making it a promising tool for motor rehabilitation. This volitional modulation induces sensorimotor cortical activation and promotes neuroplasticity by engaging new brain circuits, potentially accelerating functional recovery.
\begin{figure*}[!tb]
     \centerline{
        \begin{tabular}{c}
        \resizebox{18cm}{!}{\includegraphics{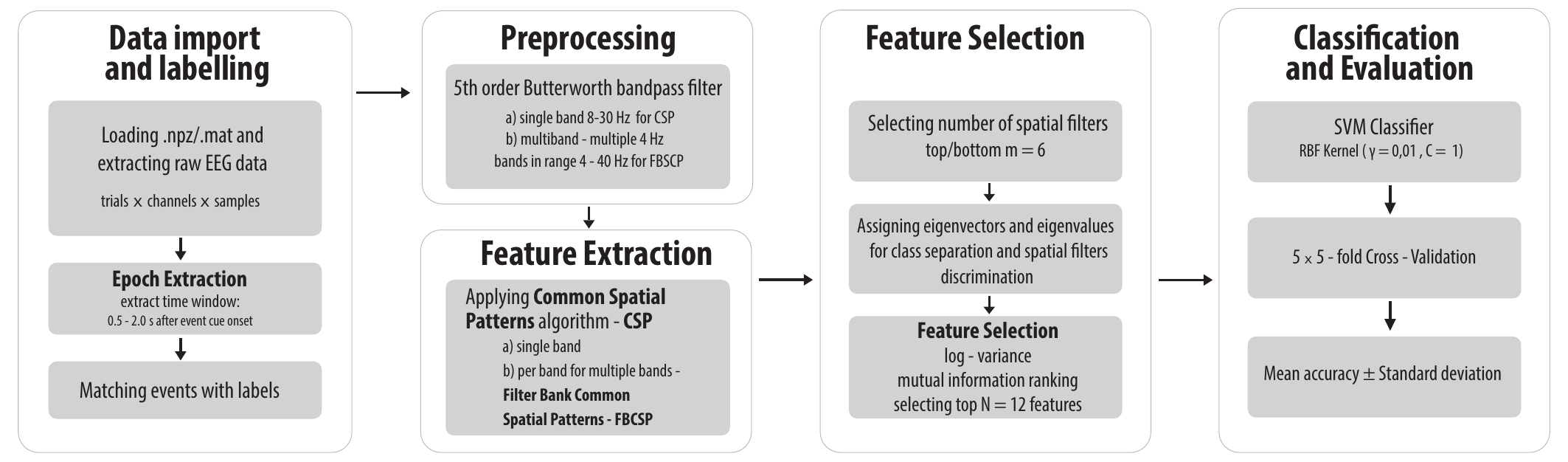}}
        \end{tabular}
    }
    \caption{A schematic representation of the MI-based BCI pipeline.}
    \label{fig:Fig_pipeline}
\end{figure*}

There is evidence that BCI-based rehabilitation can be more effective than traditional rehabilitation. However, its effectiveness largely depends on the system's reliability to detect motor imagery, which consequently depends on the subject’s ability to perform motor imagery. Usually, MI-BCI systems operate by detecting a decrease in the $\mu$ (8–12 Hz) and $\beta$ (13–30 Hz) band powers over the sensorimotor area. Left vs. right-hand MI paradigms for upper-limb rehabilitation rely on contralateral brain activation \cite{de2020effects}. However, this pattern of contralateral-lateralized activation is often disrupted in stroke patients. Studies have shown that during passive movement of the unaffected hand, brain activation typically follows the expected contralateral-lateralized pattern. In contrast, passive movement of the affected hand tends to elicit compensatory activation in the ipsilateral (i.e., contralesional) hemisphere \cite{jones2017motor, xiong2021motor}. This compensatory activation is more pronounced in patients with greater motor impairments, reflecting a loss of lateralization \cite{cirstea2000compensatory}. Such altered activation patterns can negatively affect the performance of left vs. right-hand MI paradigms. 

In this study, we investigate an alternative rehabilitation paradigm, utilizing affected-hand movement versus rest. This approach has the advantage of simplicity, as it requires the patient to perform only one type of motor imagery. 
Since MI training can gradually restore typical contralateralization by reducing compensatory ipsilateral activation \cite{xiong2021motor}, we can use personalized rehabilitation by dynamically adapting MI-BCI systems as patients progress, beginning with a movement vs. rest paradigm and changing to the more standard left vs. right-hand MI. Additionally, we explore the impact of short training sessions by dividing the data into smaller subsets.


\section{MATERIALS AND METHODS}

Figure \ref{fig:Fig_pipeline} illustrates the pipeline of the MI-based BCI, comprising the following steps: data acquisition (public datasets), preprocessing, feature extraction and selection, classification, and evaluation metrics, which are described in the following subsections.

\subsection{BCI Datasets}

The proposed approach was evaluated using two benchmark datasets: BCI Competition IV (Dataset II-a) \cite{brunner2008bci} and a publicly available dataset of 50 stroke patients \cite{liu2024eeg}. The description of each dataset is provided below.

\subsubsection{BCI Competition IV Dataset}

This dataset contains EEG data from 9 subjects performing four motor imagery tasks: left hand, right hand, both feet, and tongue. In this study, only the left and right-hand movements are used. The experiment includes two sessions conducted on different days, with 288 trials per session. 
EEG signals were recorded using 22 channels at a sampling rate of 250 Hz. For more details, see \cite{brunner2008bci}. 

\subsubsection{Stroke Dataset}

This dataset was acquired from 50 acute stroke patients performing left and right-hand motor imagery tasks.
Eight subjects were excluded due to EEG signal artifacts.
The experiment consists of 40 trials, each lasting 8 seconds, and is conducted in a single session.
EEG signals were recorded using 30 channels at a sampling rate of 500 Hz (see \cite{liu2024eeg} for details).

\subsection{Data Preprocessing}

The EEG signals were segmented into 1.5-second epochs, starting 0.5 seconds after the onset of each stimulus for left and right hand movements. For the rest condition, epochs began 0.5 seconds after the start of each trial in the BCI competition IV dataset and during the break periods in the stroke dataset. 
All epochs were bandpass filtered from 4 to 40 Hz using a Butterworth filter.

For ERD/ERS analysis, EEG signals were preprocessed using Common Average Referencing (CAR).
The EEGLAB Event-Related Spectral Perturbation (ERSP) function was used to visualize time-frequency power changes (in dB) relative to baseline.

\begin{figure*}[!tb]
     \centerline{
        \begin{tabular}{c}
        \resizebox{16.5cm}{!}{\includegraphics{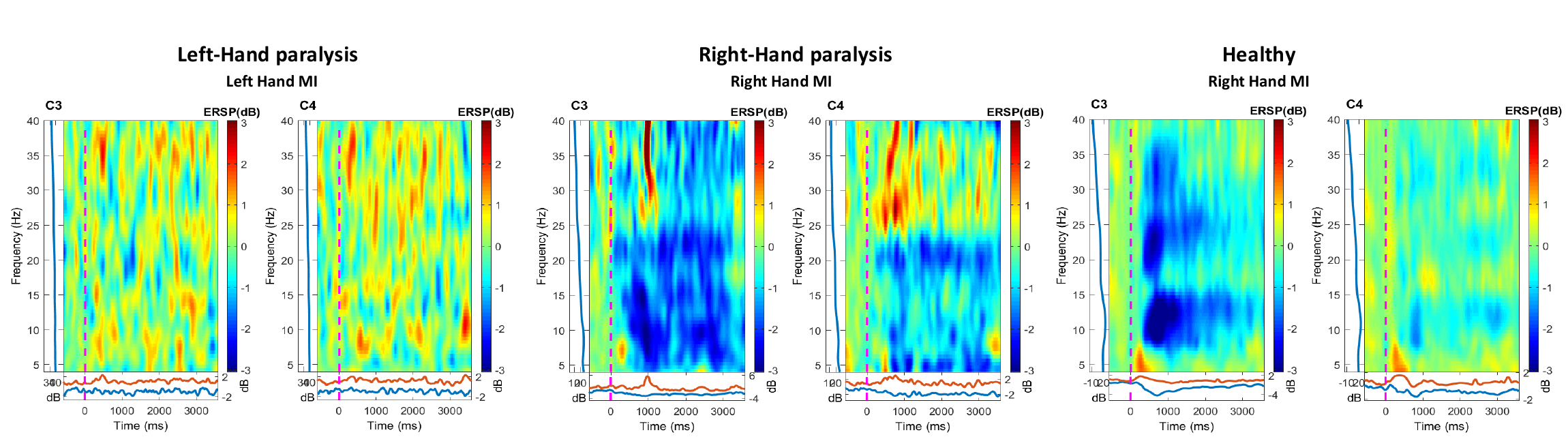}}
        \end{tabular}
    }
    \caption{ERSP spectrum of central region in healthy subjects, LHP and RHP patients.}
    \label{fig:Fig_ERD}
\end{figure*}

\begin{figure*}[!tb]
    \centerline{
        \begin{tabular}{@{\hskip 0pt}c@{\hskip 1pt}c@{\hskip 1pt}c@{\hskip 0pt}}
            \resizebox{5.5 cm}{!}{\includegraphics{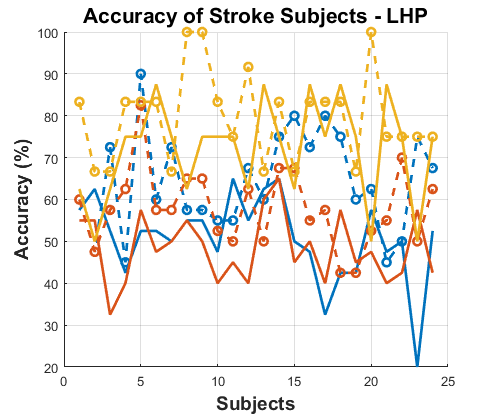}} &
            \resizebox{5.5 cm}{!}{\includegraphics{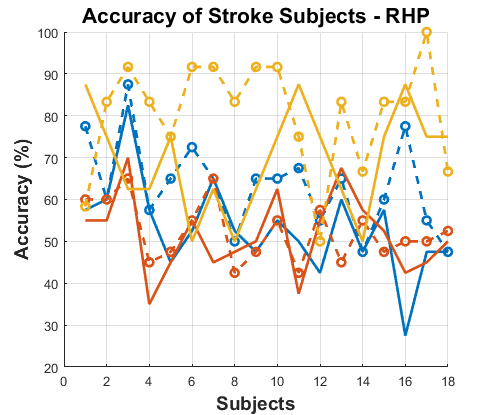}} &
            \resizebox{5.5 cm}{!}{\includegraphics{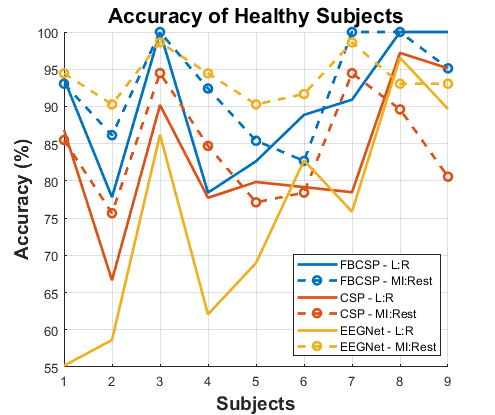}} 
        \end{tabular}
    }
    \caption{Classification performance for healthy subjects, and patients with LHP and RHP, using three different approaches. L:R refers to the classification between left and right-hand motor imagery. MI:Rest is the classification between motor imagery (left MI for LHP, right MI for RHP and healthy subjects) and rest.}
    \label{fig:Fig_class}
\end{figure*}
 
\subsection{Feature Extraction and Classification} \label{sec:CSP}

The feature extraction was performed with Common Spatial Patterns (CSP) and its extension filter bank CSP (FBCSP) \cite{ang2008filter}.
CSP is a popular spatial filtering technique used to enhance the discrimination of two different classes in motor imagery tasks. 
The goal of CSP is to find spatial filters that maximize the variance for one class while simultaneously minimizing it for the other. This is achieved by maximizing the generalized Rayleigh quotient: \( J(W)= \frac{W^TC_1W}{W^TC_2W} \), where $C_i ~~i\in{1,2} $ is the two classes’ covariance matrices.
The selected filters are the eigenvector associated with the higher and lower eigenvalues calculated by solving the generalized eigenvalue problem: \( C_2^{-1}C_1W = \lambda W \).

We employed the CSP algorithm with six spatial filters (m = 6) and a frequency band of 8–30 Hz. The FBCSP uses a filter bank from 4 to 40Hz, which contains 9 bandpass filters that covers 4Hz each. The variance of each spatial filter projection (\( \mathrm{var}_t \)) was then calculated. These values were then normalized and log-transformed as follows: \( \text{feat}_t = \log_{10} \left( \frac{\text{var}_t}{\sum_{t=1}^{m} \text{var}_t} \right) \). Then, the mutual information was used to select the 12 most informative features. Finally, these features were classified using SVM. Additionally, EEGNet, a reference deep learning approach in the context of EEG, was also employed for comparison.




\section{RESULTS AND DISCUSSION}

\subsection{Event-Related De/Synchronization Signal}

ERD and ERS are key neural markers for distinguishing right- and left-hand motor imagery. We analyzed these patterns in healthy individuals and in stroke patients with left-hand (LHP) or right-hand paralysis (RHP). Fig. \ref{fig:Fig_ERD} shows ERD/ERS results from motor cortex channels C3 (left hemisphere) and C4 (right hemisphere), commonly used in MI-based BCI systems. The results showed that, in healthy subjects, there is a clear contralateral activation: during right-hand MI, the ERD is more pronounced in C3, while during left-hand MI, there is a bilateral activation with greater activation in C4. For LHP patients, the ERSP patterns appear weaker, more variable, and diffusely distributed. In contrast, RHP patients exhibit bilateral activation, with greater activation in C3. These findings are consistent with previous results reported in the literature \cite{lee2023effects}.

\subsection{Classification Performance}

            
            
            

The proposed MI vs. rest (MI:Rest) approach was evaluated using three classification methods: CSP, FBCSP, and EEGNet, and compared to the conventional left vs. right (L:R) paradigm. The classification accuracy, obtained using 5-fold cross-validation, is presented in Fig. \ref{fig:Fig_class} for each subject of the three groups: healthy individuals, patients with LHP and RHP. For RHP and healthy subjects, the imagined movement is right-hand MI, while for LHP patients, it is left-hand MI. The classification was performed between MI and rest. The MI vs. rest approach generally showed better performance than the left vs. right approach. However, for stroke patients, accuracy varied widely, ranging from approximately 30\% in some subjects to 100\% in others.

Table \ref{tab:class_results} summarizes the classification results for the three groups and the different methods. On average, healthy subjects achieved higher accuracy than stroke patients. MI vs. rest condition showed superior classification performance compared to the left vs. right condition. Paired sample t-tests were conducted between L:Rest and L:R, and between R:Rest and L:R.
The * in Table \ref{tab:class_results} indicates that the difference is statistically significant ($p < 0.05$). EEGNet showed significant improvements in both groups, while FBCSP was significant in all conditions except R:Rest in the healthy group. 
These results suggest that the proposed approach enhances MI classification in stroke patients, with potential applicability in BCI-based neurorehabilitation strategies. 

The results in Table \ref{tab:class_results} align with state-of-the-art findings, as reported in the review \cite{vavoulis2023review}, classification accuracies for healthy and stroke patients range from 62\% to 95\%, with an average of 77\%.
Task gamification is a promising approach to further enhance engagement and performance.

\begin{figure}[!tb]
     \centerline{
        \begin{tabular}{c}
        \resizebox{8.5cm}{!}{\includegraphics{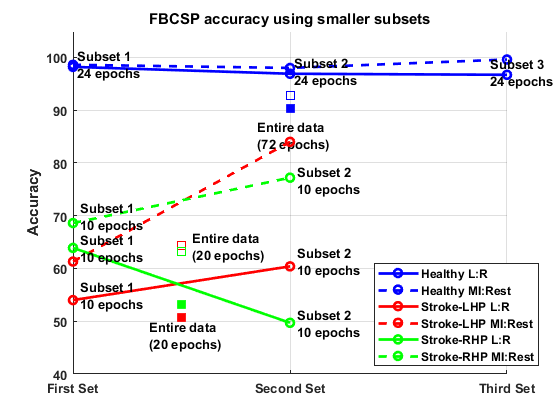}}
        \end{tabular}
    }
    \caption{Classification accuracy of the FBCSP algorithm using smaller subsets of the original datasets (three subsets for the BCI Competition IV dataset and two for the stroke dataset).}
    \label{fig:Fig_trial}
\end{figure}

We also analyzed the impact of reducing the duration of the training session. In this experiment, we divided the BCI competition IV dataset into three sets (24 epochs each) and the stroke dataset into two sets (10 epochs each). We then performed 5-fold cross-validation using the FBCSP algorithm.
As shown in Fig. \ref{fig:Fig_trial}, most of the data subsets achieved higher accuracy than using the entire dataset at once.
For stroke subjects, the mean classification accuracies for L:R, L:Rest, and R:Rest were 57.2\%, 72.6\%, and 72.2\% for the LHP and 56.8\%, 67.5\%, and 72.9\% for RHP, respectively. For healthy subjects, the mean accuracy was greater than 96.7\% for all sets. 
These results suggest that data variability may hinder MI performance. Therefore, many short training sessions or adapting classification models with new samples may enhance MI performance during rehabilitation.

\begin{table}[!tb]
  \centering
  \caption{A classification accuracy summary using different methods tested on three groups. * indicates a statistically significant higher value than L:R ($p < 0.05$).}
    \begin{tabular}{lrrrrrr}
    \toprule
          & \multicolumn{6}{c}{Healthy Subjects} \\
    \midrule
          & \multicolumn{2}{c}{FBCSP} & \multicolumn{2}{c}{CSP} & \multicolumn{2}{c}{EEGNet} \\
    \midrule
          & \multicolumn{1}{l}{Mean} & \multicolumn{1}{l}{Std} & \multicolumn{1}{l}{Mean} & \multicolumn{1}{l}{Std} & \multicolumn{1}{l}{Mean} & \multicolumn{1}{l}{Std} \\
    \midrule
    L:R   & 90.3  & 9.0   & 83.5  & 9.7   & 75.1~  & 14.7 \\
    L:Rest & \text{94.7*}  & 5.4   & 87.9    & 5.1   & \text{92.6*}   & 3.8 \\
    R:Rest & 92.8  & 6.7   & 84.5  & 7.2   & \text{93.8*}   & 3.1 \\
    \midrule
          & \multicolumn{6}{c}{Stroke Subjects Left Hand Affected} \\
    \midrule
    L:R   & 50.7~  & 10.1   & 48.3  & 8.1   & 71.9~  & 11.8 \\
    L:Rest & \text{64.3*}    & 12.0  & \text{58.0*}  & 9.2   & \text{79.9*}   & 10.7 \\
    R:Rest & \text{65.8*} & 10.2  & \text{57.7*}  & 9.4   & \text{79.9*}   & 12.5 \\
    \midrule
          & \multicolumn{6}{c}{Stroke Subjects Right Hand Affected} \\
    \midrule
    L:R   & 53.1~  & 11.2    & 51.5  & 9.4  & 69.4~  & 12.3 \\
    L:Rest & \text{61.9*}   & 15.1    & 55.3  & 8.5  &  \text{78.7*}   & 10.8 \\
    R:Rest & \text{63.3*}   & 10.8  & 52.4    & 7.1   &  \text{80.6*}   & 13.1 \\ \hline
    \end{tabular}%
  \label{tab:class_results}%
\end{table}%
\vspace{-1pt} 

\section{CONCLUSIONS}

Two strategies were explored to improve MI-based BCI protocols in stroke rehabilitation. The first strategy involved classifying affected hand movement versus rest, rather than using the conventional left vs. right approach. This approach achieved better classification performance in stroke patients. The second strategy was to reduce the length of training sessions. The results showed that using smaller subsets of data improves MI performance compared to using the entire dataset at once. This suggests that multiple short training sessions or adaptive training protocols may enhance MI classification. Overall, these findings highlight the importance of both task design and training protocol in optimizing BCI-based rehabilitation for stroke patients.

\addtolength{\textheight}{-12cm}   




\bibliographystyle{IEEEtran}

\bibliography{IEEEexample}

\begin{thebibliography}{10}
\providecommand{\url}[1]{#1}
\csname url@rmstyle\endcsname
\providecommand{\newblock}{\relax}
\providecommand{\bibinfo}[2]{#2}
\providecommand\BIBentrySTDinterwordspacing{\spaceskip=0pt\relax}
\providecommand\BIBentryALTinterwordstretchfactor{4}
\providecommand\BIBentryALTinterwordspacing{\spaceskip=\fontdimen2\font plus
\BIBentryALTinterwordstretchfactor\fontdimen3\font minus \fontdimen4\font\relax}
\providecommand\BIBforeignlanguage[2]{{%
\expandafter\ifx\csname l@#1\endcsname\relax
\typeout{** WARNING: IEEEtran.bst: No hyphenation pattern has been}%
\typeout{** loaded for the language `#1'. Using the pattern for}%
\typeout{** the default language instead.}%
\else
\language=\csname l@#1\endcsname
\fi
#2}}

\bibitem{feigin2022world}
V.~L. Feigin, M.~Brainin, B.~Norrving, S.~Martins, R.~L. Sacco, W.~Hacke, M.~Fisher, J.~Pandian, and P.~Lindsay, ``World stroke organization (wso): global stroke fact sheet 2022,'' \emph{International Journal of Stroke}, vol.~17, no.~1, pp. 18--29, 2022.

\bibitem{emerson2018combined}
J.~R. Emerson, J.~A. Binks, M.~W. Scott, R.~P. Kenny, and D.~L. Eaves, ``Combined action observation and motor imagery therapy: a novel method for post-stroke motor rehabilitation,'' \emph{AIMS neuroscience}, vol.~5, no.~4, p. 236, 2018.

\bibitem{wolpaw2013brain}
J.~R. Wolpaw, ``Brain--computer interfaces,'' in \emph{Handbook of clinical neurology}.\hskip 1em plus 0.5em minus 0.4em\relax Elsevier, 2013, vol. 110, pp. 67--74.

\bibitem{neuper2014b}
C.~Neuper and B.~Allison, ``The b of bcis: Neurofeedback principles and how they can yield clearer brain signals,'' \emph{Different psychological perspectives on cognitive processes: current research trends in Alps-Adria region}, pp. 133--153, 2014.

\bibitem{baniqued2021brain}
P.~D.~E. Baniqued, E.~C. Stanyer, M.~Awais, A.~Alazmani, A.~E. Jackson, M.~A. Mon-Williams, F.~Mushtaq, and R.~J. Holt, ``Brain--computer interface robotics for hand rehabilitation after stroke: a systematic review,'' \emph{Journal of neuroengineering and rehabilitation}, vol.~18, pp. 1--25, 2021.

\bibitem{de2020effects}
M.~de~Castro-Cros, M.~Sebastian-Romagosa, J.~Rodr{\'\i}guez-Serrano, E.~Opisso, M.~Ochoa, R.~Ortner, C.~Guger, and D.~Tost, ``Effects of gamification in bci functional rehabilitation,'' \emph{Frontiers in neuroscience}, vol.~14, p. 882, 2020.

\bibitem{jones2017motor}
T.~A. Jones, ``Motor compensation and its effects on neural reorganization after stroke,'' \emph{Nature Reviews Neuroscience}, vol.~18, no.~5, pp. 267--280, 2017.

\bibitem{xiong2021motor}
X.~Xiong, H.~Wang, X.~Wang, L.~Sun, and X.~Guo, ``Motor imagery training reduces contralesional compensation in stroke patients with moderate to severe upper limb impairment,'' in \emph{2021 10th International IEEE/EMBS Conf. on Neural Eng. (NER)}.\hskip 1em plus 0.5em minus 0.4em\relax IEEE, 2021, pp. 876--879.

\bibitem{cirstea2000compensatory}
M.~Cirstea and M.~F. Levin, ``Compensatory strategies for reaching in stroke,'' \emph{Brain}, vol. 123, no.~5, pp. 940--953, 2000.

\bibitem{brunner2008bci}
C.~Brunner, R.~Leeb, G.~M{\"u}ller-Putz, A.~Schl{\"o}gl, and G.~Pfurtscheller, ``Bci competition 2008--graz data set a,'' \emph{Institute for knowledge discovery (laboratory of brain-computer interfaces), Graz University of Technology}, vol.~16, no. 1-6, p.~1, 2008.

\bibitem{liu2024eeg}
H.~Liu, P.~Wei, H.~Wang, X.~Lv, W.~Duan, M.~Li, Y.~Zhao, Q.~Wang, X.~Chen, G.~Shi, \emph{et~al.}, ``An eeg motor imagery dataset for brain computer interface in acute stroke patients,'' \emph{Scientific Data}, vol.~11, no.~1, p. 131, 2024.

\bibitem{ang2008filter}
K.~K. Ang, Z.~Y. Chin, H.~Zhang, and C.~Guan, ``Filter bank common spatial pattern (fbcsp) in brain-computer interface,'' in \emph{2008 IEEE international joint conference on neural networks (IEEE world congress on computational intelligence)}.\hskip 1em plus 0.5em minus 0.4em\relax IEEE, 2008, pp. 2390--2397.

\bibitem{lee2023effects}
S.~Lee, H.~Kim, J.~B. Kim, and D.-J. Kim, ``Effects of altered functional connectivity on motor imagery brain--computer interfaces based on the laterality of paralysis in hemiplegia patients,'' \emph{Computers in Biology and Medicine}, vol. 166, p. 107435, 2023.

\bibitem{vavoulis2023review}
A.~Vavoulis, P.~Figueiredo, and A.~Vourvopoulos, ``A review of online classification performance in motor imagery-based brain-computer interfaces for stroke neurorehabilitation,'' \emph{Signals}, vol.~4, no.~1, pp. 73--86, 2023.

\end{thebibliography}

\end{document}